# Relativistic electrodynamics as an extrapolation of relativistic kinematics

## Bernhard Rothenstein[1] and Stefan Popescu[2]

1) Politehnica University of Timisoara, Physics Department,
Timisoara, Romania brothenstein@gmail.com
2) Siemens AG, Erlangen, Germany stefan.popescu@siemens.com

**Abstract.** *After having identified all the possible relationships between the electric field and the magnetic field in a given inertial reference frame we derive the transformation equations for the components of these fields. Special relativity is involved via the addition law of parallel speeds or via the Lorentz transformations for the space-time coordinates of the same event. Electricity and magnetism are involved via Gauss's and Ampere's laws. In this way we avoid the transformation equations for the Lorenz force components which are used in most derivations of the transformation equations for **E** and **B** fields.*

## 1. Introduction

Consider an empty space referred to an inertial reference frame I(XYZO). It hosts distributed electric charge at rest or in a state of uniform motion. Experiment tells us that a test electric charge $q$ passing through a point M(x,y,z) with velocity **u** experiences a velocity independent electric force

$$\mathbf{F}_e = q\mathbf{E} \tag{1}$$

and a velocity dependent force

$$\mathbf{F}_m = q\mathbf{u} \times \mathbf{B}. \tag{2}$$

The electric field **E** at the point where the probe charge is located is defined by taking the limit in equation (1) with the probe charge becoming infinite small:

$$\mathbf{E} = \lim_{q \to 0} \frac{\mathbf{F}_e}{q}. \tag{3}$$

Equation (2) tells us that the magnitude of the magnetic force

$$F_m = quB\sin\theta \tag{4}$$

depends on the angle $\theta$ between the vectors **u** and **B** presenting a maximum for $\theta = \pi/2$ equal to

$$F_{m,\max} = quB \tag{5}$$

It enables us to define the magnitude of the magnetic field at the same point in space by taking the limit in equation (5) with the probe charge becoming infinite small:

$$B = \lim_{q \to 0} \frac{F_{m,\max}}{uq}. \tag{6}$$

Because $\mathbf{F_e}$ and $\mathbf{F_m}$ could have different orientations in space the Lorentz force (the sum of the two forces) is presented in vector form as[1]

$$\mathbf{F} = \mathbf{F}_e + \mathbf{F}_m = q(\mathbf{E} + \mathbf{u} \times \mathbf{B}) \qquad (7)$$

Rosser[1] presents three different ways to derive the transformation equations for the normal components of **E** and **B** involving partial derivatives of the involved fields and the transformation equations for the components of force. Our derivation is based on the kinematics roots of electrodynamics, involving the Gauss's and Ampere's law and the Lorentz transformations for the space-time coordinates of the same event or alternatively the addition law of parallel speeds.

## 2. A scenario which leads to the possible relationships between E and B

The Lorentz transformations for the electric and the magnetic fields are performed in very different ways. The difference consists in the way in which special relativity is involved and in the way in which **E** and **B** are supposed to be generated. The scenario presented in Figure 1 involves an infinite long wire, uniformly charged with a linear density of electric charge $\lambda_0$ (proper magnitude). This wire is located along the OX axis and moves with constant velocity $u_{1,x}$ in its positive direction.

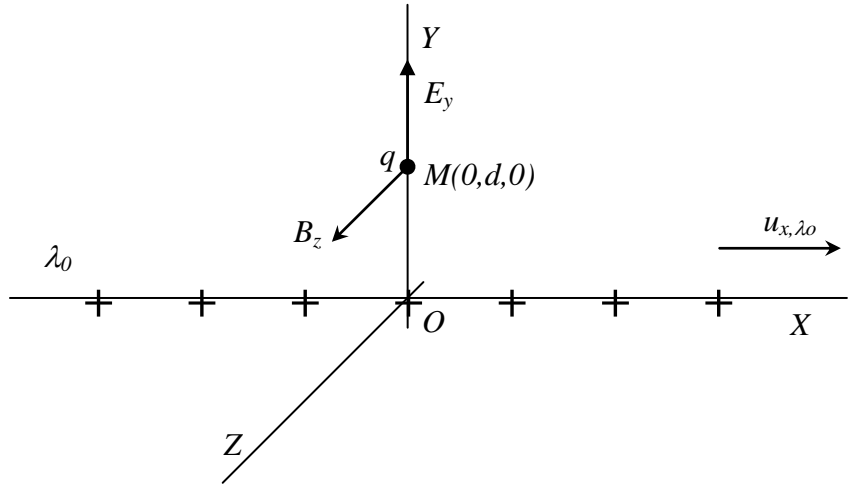

*Figure 1. Scenario for deriving the relationship between the velocity independent electric field and the magnetic field generated by an uniformly distributed linear density of charge moving with speed $u_{x,\lambda_0}$*

As a result of the invariance of the electric charge, of Gauss' theorem and of the relativistic length contraction effect this distribution of electric charge generates at a point M(x=0,y=d,z=0) an electric field

$$E_y = \frac{\lambda_0}{2\pi\varepsilon_0 d \sqrt{1 - \frac{u_{x,\lambda_0}^2}{c^2}}} \qquad (8)$$

that shows in the positive direction of the OY axis. The moving distribution of electric charge is equivalent with an electric current given by

$$I = \frac{\lambda_0 u_{x,\lambda_0}}{\sqrt{1 - \frac{u_{x,\lambda_0}^2}{c^2}}} \tag{9}$$

that generates at the point M(0,d,0) a magnetic field

$$B_z = \mu_0 \frac{\lambda_0 u_{x,\lambda_0}}{2\pi d \sqrt{1 - \frac{u_{x,\lambda_0}^2}{c^2}}} \tag{10}$$

showing in the positive direction of the OX axis. Taking into account that $\varepsilon_0$ and $\mu_0$ are related by

$$c^2 = (\varepsilon_0 \mu_0)^{-1} \tag{11}$$

and combining (8) and (10) we obtain that $E_y$ and $B_z$ are related by

$$B_z = \frac{u_{x,\lambda_0} E_y}{c^2} \tag{12}$$

This equation represents a first possible relationship between the two components of the electric and magnetic fields both generated by the same linear distribution of electric charges.

A point-like test charge $q$ at rest in I and located at the point M(0,d,0) is acted upon by a single "electrostatic" force

$$F_{y,e} = qE_y. \tag{13}$$

So far the magnetic field $B_z$ is supposed to exist but it doesn't act on the stationary test charge.

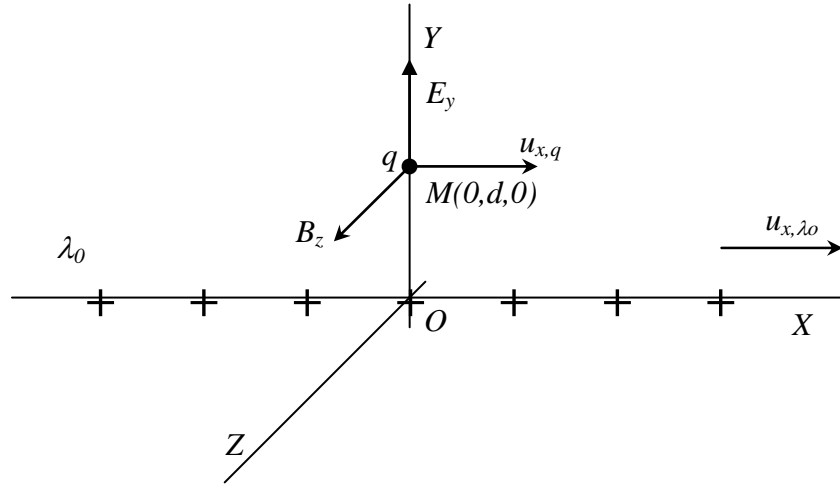

*Figure 2*. *The same scenario as in Figure 1 involving a test charge q that moves with speed $u_{x,q}$ at a distance d from the linear distribution of electric charges.*

Figure 2 depicts the same scenario considering that the test charge *q* moves with constant speed $u_{x,q}$. Under such conditions it is acted upon by the electric force $F_{y,e}=qE_y$ (defined above) and by a supplementary force
$$F_{y,m} = -qu_{x,q}B_z. \tag{14}$$
The resultant force acting upon the test charge is given by
$$F_y = F_{y,e} + F_{y,m} = q(E_{y,e} - u_{x,q}B) \tag{15}$$
In the particular case when the particle velocity fulfills the condition
$$E_{y,e} = u_{x,q}B_z \tag{16}$$
the test charge will not be deviated from its direction of motion and such a "device" acts as a velocity filter. We find a similar situation in the case when the test charge moves within a space that hosts two orthogonal electric and magnetic fields as in the case of the Wien's velocity filter.[2] In this case the electric and the magnetic fields are generated by different sources. Equation (16) is a possible relationship between the normal components of the electric and the magnetic field.

We further notice that equation (16) is similar to $E_y = cB_z$ the later corresponding to the relation between the orthogonal field components of a plane electromagnetic wave traveling in vacuum at group velocity *c* along the OX axis. Without analyzing this aspect in further detail we observe that any charge moving at speed < *c* within the electromagnetic wave will be deviated. Other formulated: if the electromagnetic wave would transport any kind of charge in vacuum then it should travel at light speed in order to maintain direction with the wave. Wave propagation through matter with group velocity less that *c* may change this situation.

**3. Two simple derivations of the transformation equations for the normal components of the electric and magnetic field**

**3.1. Using the Lorentz transformation for the space-time coordinates of the same event**
Consider the same scenario above this time as seen from another reference frame I' that moves with constant velocity *V* relative to frame I along its OX axis. In accordance with the first relativistic postulate (12) and (16) read now
$$B'_z = \frac{u'_x E'_y}{c^2} \tag{17}$$
and respectively
$$E'_y = u'_x B'_z \tag{18}$$

We remind at this point some results known from relativistic kinematics. Consider a particle that passes through the origins O and O' at the origin of time in the two involved inertial reference frames (t=t'=0) with speed $u_x$ and $u'_x$ relative to I and I' respectively. After a given time of motion t it travels a distance x related to t by

$$x = u_x t \tag{19}$$

that reads in I'

$$x' = u'_x t'. \tag{20}$$

The Lorentz transformations for the space-time coordinates of the same event tell us that

$$x = x' \frac{1 + V/u'_x}{\sqrt{1 - V^2/c^2}} \tag{21}$$

and that

$$t = t' \frac{1 + V u'_x/c^2}{\sqrt{1 - V^2/c^2}}. \tag{22}$$

Comparing (16) and (18) with (19) and (20) we consider that $E_y(E'_y)$ should transform as space coordinates do with $B_z(B'_z)$ transforming as time coordinates do. The consequence is that the transformation equations for space-time coordinates extrapolated to the electric and magnetic fields should hold true, thus we shall have:

$$B_z = B'_z \frac{1 + V/u'_x}{\sqrt{1 - V^2/c^2}} = \frac{B'_z + V E'_y/c^2}{\sqrt{1 - V^2/c^2}} \tag{23}$$

and

$$E_y = E'_y \frac{1 + V u'_x/c^2}{\sqrt{1 - V^2/c^2}}. \tag{24}$$

Alternatively comparing (12) and (17) with (19) and (20) when the particle speed is $u_x/c^2$ we consider that $E_y/c^2$ and $E'_y/c^2$ should transform as time coordinates do with $B_z(B'_z)$ transforming as space coordinates do. The consequence is that we recover the same transformation equations derived above.

Considering the fields $E_z$ and $B_y$ generated by the same distribution of electric charges at a point M(0,0,d) located on the OZ axis as shown in figure 3.

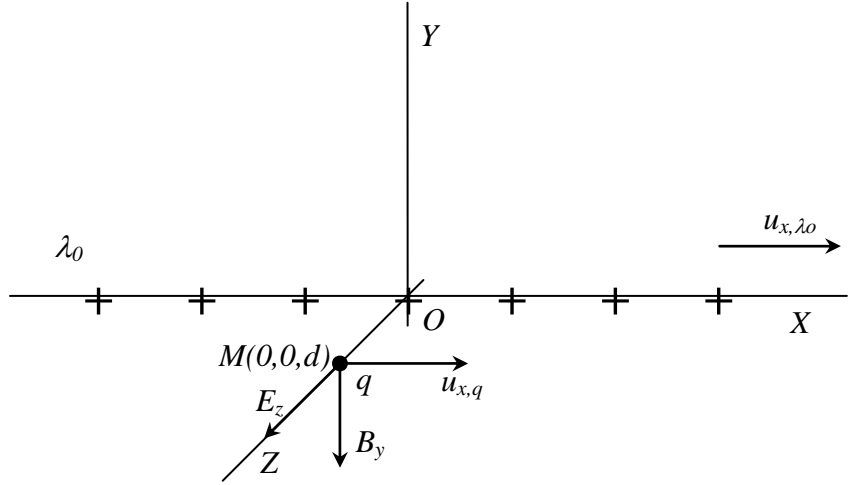

*Figure 3. The electric and the magnetic field generated by the same linear distribution of electric charges at a point M(0,0,d) located on the OZ axis.*

In this arrangement the electric field $E_z$ shows in the positive direction of the OZ axis with the magnetic field showing in the negative direction of the OY axis. A simple exercise shows that the corresponding transformation equations are

$$E_z = \frac{E'_z - VB'_y}{\sqrt{1 - V^2/c^2}} \qquad (25)$$

and

$$B_y = \frac{B'_y - VE'_z/c^2}{\sqrt{1 - V^2/c^2}}. \qquad (26)$$

### 3.2. Using the addition law of parallel velocities

Consider the scenario presented in Figure 1 from an inertial reference frame $I^0$ relative to which the linear distribution of electric charges is in a state of rest. Observers of this reference frame detect the presence of the electric field

$$E_y^0 = \frac{\lambda_0}{2\pi\varepsilon_0 d} \qquad (27)$$

but no magnetic field. The reference frame $I^0$ moves with speeds $u_x$ relative to I and $u'_x$ relative to I'. In accordance with the addition law of parallel speeds we have

$$u_x = \frac{u'_x + V}{1 + \frac{u'_x V}{c^2}}. \qquad (28)$$

Observers from I detect the presence of an electric field

$$E_y = \frac{E_y^0}{\sqrt{1 - \frac{u_x^2}{c^2}}} \qquad (29)$$

whereas the observers from I' detect the presence of an electric field

$$E'_y = \frac{E^0_y}{\sqrt{1-\dfrac{u'^2_x}{c^2}}}. \tag{30}$$

Expressing the right side of (29) as a function of $u'_x$ via (30) and taking into account (37) we recover the same transformation equations derived in **3.1.**

In order to derive the transformation equations for the $E_z(E'_z)$ and $B_y(B'_y)$ consider the scenario presented in Figure 1 at the point $M_1(x=0,y=0,z=d)$. When detected from I the linear distribution of electric charges generates the electric field

$$E_z = \frac{\lambda_0}{2\pi\varepsilon_0 d\sqrt{1-\dfrac{u^2_x}{c^2}}} \tag{31}$$

and the magnetic field

$$B_z = -\frac{u_x E_z}{c^2}. \tag{32}$$

Following one of the approaches presented above we can derive the transformation equations for the components of the **E** and **B** fields.

### 4. Conclusions

The derivations of the transformation equations for the components of **E** and **B** presented above illustrate the fact that they do not involve Maxwell's equations, being based on equations derived in relativistic kinematics. The derivations are simple and transparent as compared with those we find in the current literature of the subject[1].